\documentclass{nature}
\usepackage[T1]{fontenc}
\usepackage[utf8]{inputenc}
\usepackage[T1]{fontenc}
\usepackage{float}
\usepackage{amsmath}
\usepackage {mathtools}
\usepackage{indentfirst}
\usepackage{textcomp}
\usepackage{color}
\usepackage{soul}
\usepackage{ulem}
\usepackage{siunitx}
\usepackage{amssymb}
\usepackage{graphicx}
\usepackage{caption}
\usepackage{authblk}
\usepackage{array}
\usepackage{booktabs,multirow}
\usepackage{microtype,physics} 
\usepackage{url}

\usepackage[version=4]{mhchem}

\usepackage{xr}
\makeatletter
\newcommand*{\addFileDependency}[1]{
  \typeout{(#1)}
  \@addtofilelist{#1}
  \IfFileExists{#1}{}{\typeout{No file #1.}}
}
\makeatother

\title{Towards real-time additive-free dopamine detection at $10^{-8}$ mM with hardware accelerated platform integrated on camera}

\author{N. Li$^{1\dagger}$, Q. Wang$^{1\dagger}$, Z. He$^1$, A. Burguete-Lopez$^1$, F. Xiang$^1$ and A. Fratalocchi$^{1*}$}

\begin{document}
\maketitle

\begin{affiliations}
 \item PRIMALIGHT, Faculty of Electrical Engineering, King Abdullah University of Science and Technology (KAUST), Thuwal 23955-6900, Saudi Arabia.\\ 
 \textsuperscript{\textdagger}first authors with equal contribution
\end{affiliations}

\begin{abstract}
Tracing physiological neurotransmitters such as dopamine (DA) with detection limits down to  $\mathrm{1\times10^{-8}}$ mM is a critical goal in neuroscience for studying brain functions and progressing the understanding of cerebral disease. Addressing this problem requires enhancing the current state-of-the-art additive-free electrochemical workstation methods by over two orders of magnitude. In this work, we implement an ultra-sensitive, additive-free platform exploiting suitably engineered light-scattering membranes and optical accelerators integrated into commercial vision cameras, reporting real-time detection of DA in uric and ascorbic acid below the concentration of $\mathrm{10^{-8}}$ mM. These performances improve the current best technology by over two orders of magnitude in resolution while providing continuous, real-time detection at video rates. This technology also upgrades the bulk form factor of an electrochemical workstation with an imaging camera's compact and portable footprint. The optical accelerator implemented in this work is universal and trainable to detect a wide range of biological analytes. This technology's wide adoption could help enable early disease detection and personalized treatment adjustments while improving the management of neurological, mental, and immune-related conditions.
\end{abstract}

\section*{Introduction}
\noindent Tracing analyte concentrations at ultra-low quantities is essential across various scientific areas, including security~\cite{tumay2021small,to2020recent, dai2020ultrasensitive},  medicine~\cite{liu2021ultrasensitivea,kang2021ultrasensitive,mejia-salazar2018plasmonic,sun2020skin,feng2023ultrasensitive}, food safety~\cite{ha2017ultra}, and environmental protection~\cite{zhang2020ultrasensitive, wan2019cascaded}.
In disease diagnosis, ultrasensitive detection enables tracking subtle changes of biomarker concentration, which indicate early onset or progression of debilitating conditions~\cite{sandi2021neurodegeneration}. Dopamine (DA) is a fundamental neurotransmitter biomarker whose microscopic change within the $\mathrm{10^{-7}}$ mM range~\cite{sharma2019dopamine,zbroch2013circulating,shine2019dopamine,howes2024schizophrenia} reveals various neurological disorders, including Parkinson's~\cite{xu2002dopamine,lotharius2002pathogenesis}, Alzheimer's~\cite{martorana2009dopamine}, depression~\cite{he2023prosaposin}, schizophrenia~\cite{okubo1997decreased,rolls2008computational}, and attention deficit hyperactivity disorder (ADHD)~\cite{demontis2023genome}. Practical strategies for ultrasensitive detection of DA require identifying the analyte in interference media such as uric acid (UA) and ascorbic acid (AA) without any additive incorporation of enzymes, polymers, aptamers, etc., which introduce impurities and  instabilities~\cite{wang2024dopamine,liu2021biosensors}.\\
The state-of-the-art in additive-free biosensors reach the limit of $\mathrm{5\times10^{-6}}$~mM with \ce{Mn-MoS2}/PGS sensor ~\cite{lei2020singleatom}. These biosensors do not work in real-time and necessitate measurement times from tens of seconds to several minutes~\cite{alothman2010simultaneous,liu2018electrochemical,ly2006detection}, while requiring an enhancement of more than two orders of magnitude to reach the range between $10^{-8}$~mM to $10^{-6}$~mM characteristic of a majority of brain disorders~\cite{orhan2023csf,goldstein2012cerebrospinal}.
Devising pathways to enhance these performances is challenging due to background noise interference with the signal that, at ultra-low sensitivities, yields variation of measurement results across different manufactured devices~\cite{cui2023highly,tiwari2016engineered}, making it difficult to deploy an efficient decision-making strategy that minimizes human bias~\cite{zhang2024discrimination,masson2023machine}. Addressing these issues provides a significant opportunity in research, promising early diagnosis and low-cost routine monitoring that could help progress against elusive disorders affecting tens of millions of people worldwide.\\
This work implements an ultra-sensitive detection platform leveraging artificial intelligence (AI) hardware, incorporating data acquisition and feature extraction into a single process. The hardware accelerator integrates with commercially available monochrome cameras, retaining their footprint, portability, and video rate processing speed. In experimental tests, the system reports DA detection at a concentration of $\mathrm{10^{-8}}$ mM with interference media, including UA, AA, and phosphate buffer solution (PBS). Compared to current AI-based DA sensing approaches~\cite{de2023empowering,rao2024construction,pal2023paper}, the proposed platform enhances the detection limit of DA by six orders of magnitude~\cite{rao2024construction}. When compared against existing additive-free DA detection technologies, the proposed platform enhances the state-of-the-art detection sensitivity by two orders of magnitude~\cite{lei2020singleatom} while providing a fully integrated platform that is massively deployable at the point-of-care.

\section*{Results}
\begin{figure*}
\centering
\includegraphics[width=1\textwidth]{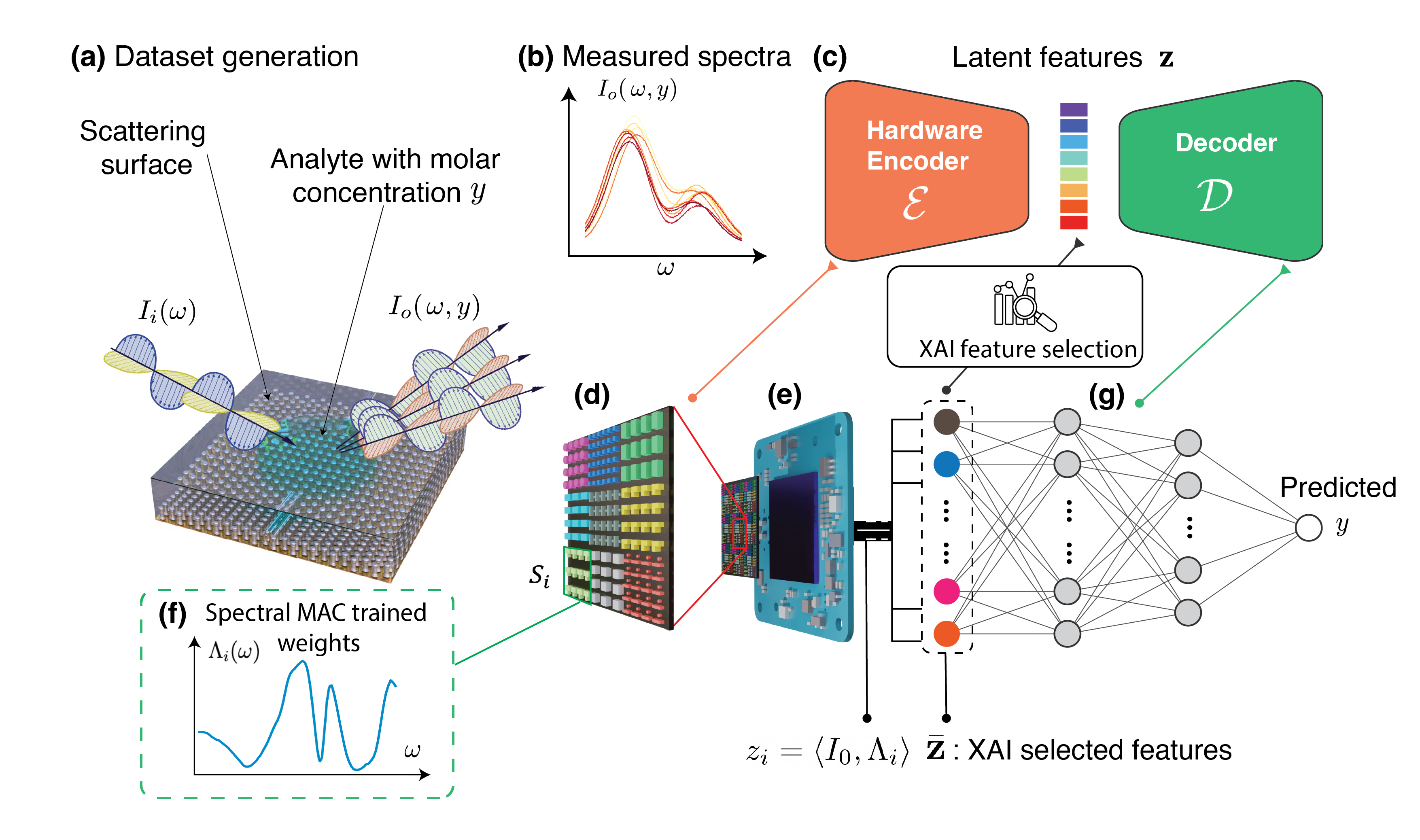}
\caption{\textbf{Schematics of the general sensing platform} (a) Spectral dataset generation using a scattering surface with broadband illumination. (b) Example of a generated spectral dataset for a varying concentration of analyte. (c) Encoder-decoder unsupervised training model with explainable-AI (XAI) interface for selecting latent features. (d) Implementation design of hardware encoder with sub-pixels of inverse-designed nanoresonators. (e) Integration of the hardware encoder into a CMOS camera sensor. (f) Software post-processing network processing the extracted latent features into the analyte concentration. (g) Each sub-pixel encoder's trained spectral transmission response implements the multiply and accumulate (MAC) neural network operation required to extract the selected latent space features.
\label{frame}
}
\end{figure*}

\begin{figure*}
\centering
\includegraphics[width=\textwidth]{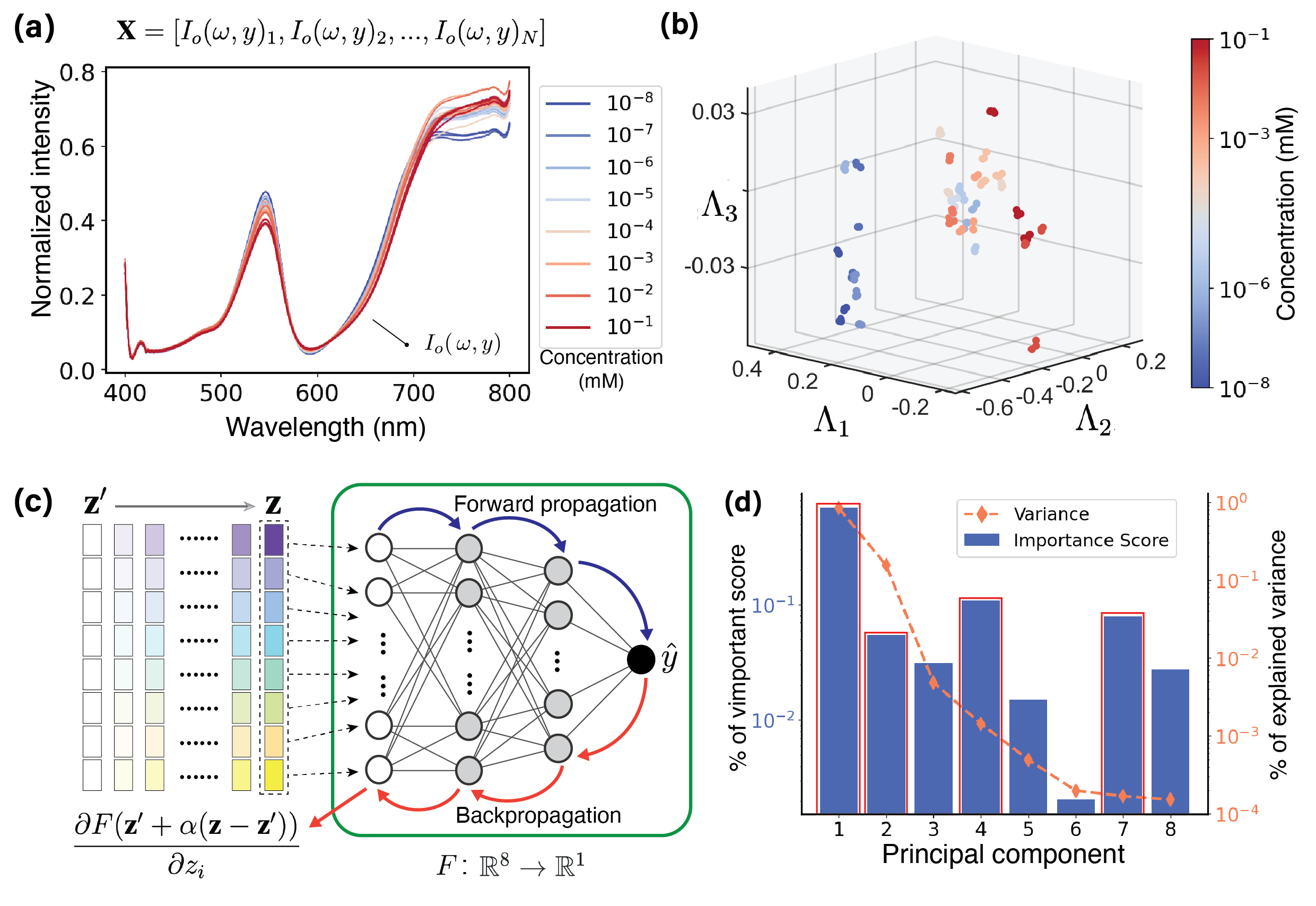}
\caption{\textbf{Design workflow of the optical hardware encoder.} (a) Example of a dataset comprising scattered spectra from DA in PBS. (b) The dataset projected along the first three principal components extracted from the data. (c) Exaplainable-AI integrated gradient method for extracting the essential features in the latent space for predicting the analyte concentration $y$. (d) Principal component's importance score vs. explained variance. Components 1-3 and 8 are the most important in computing the concentration $y$.
\label{design}
}
\end{figure*}

\begin{figure*}
\centering
\includegraphics[width=0.99\textwidth]{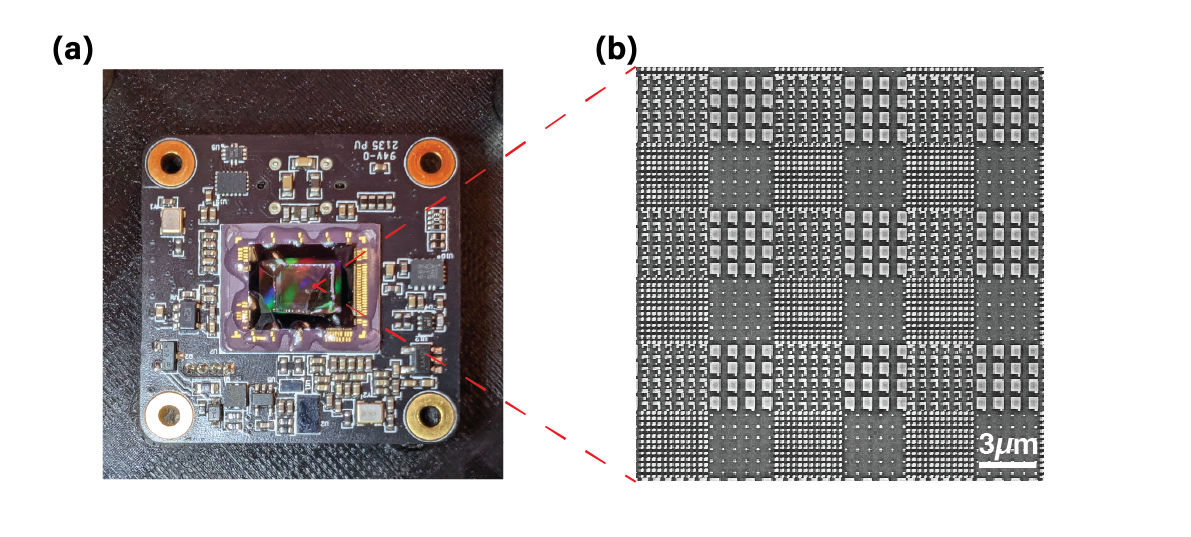}
\caption{\textbf{Hardware implementation on the camera chip of XAI-selected encoders} (a) Integration of the hardware encoders on a CMOS camera sensor. (b) SEM image of the fabricated encoder nanostructures. 
\label{integration}
}
\end{figure*}

\begin{figure*}
\centering
\includegraphics[width=0.99\textwidth]{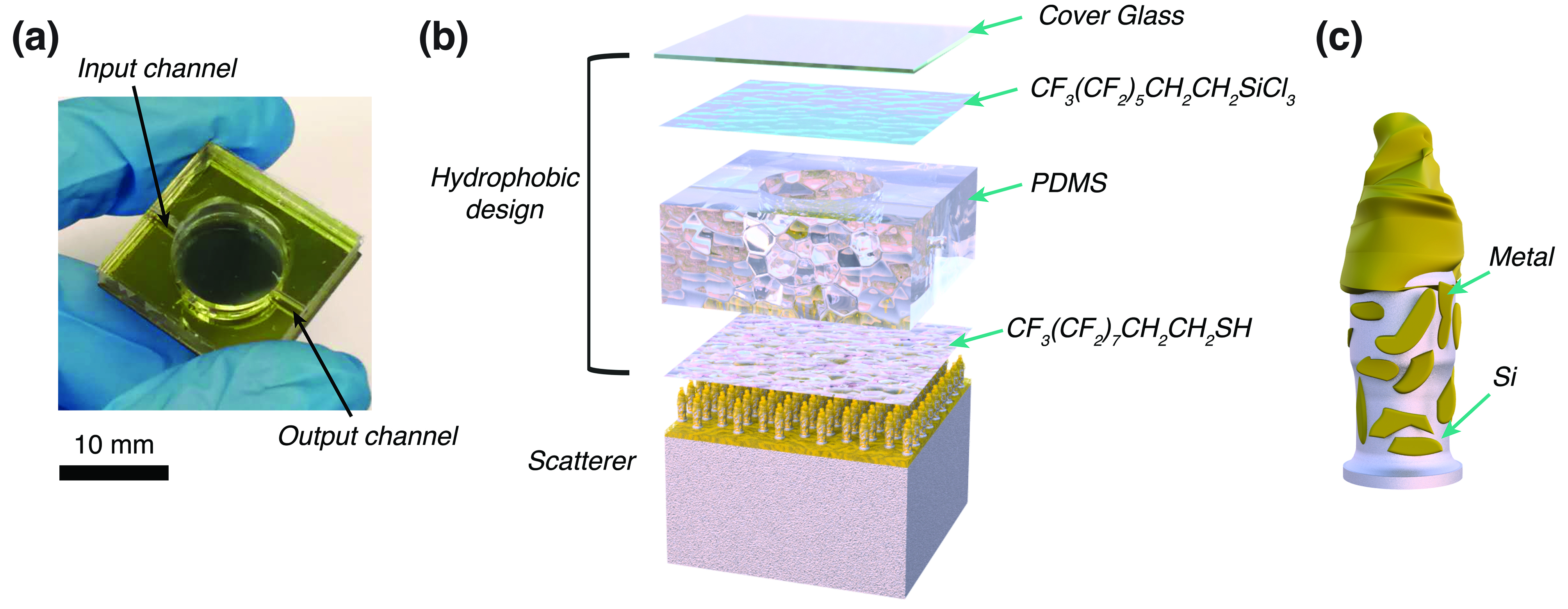}
\caption{\textbf{Implementation of the scattering device.} 
(a) Optical image of an experimentally assembled scattering device comprising a reusable microfluidic cell with input and output channels. (b) Schematic assembly of the microfluidic device composed of the scatterer, PFDT, PDMS, FDTS, and the cover glass. (c) Schematic of one scattering pillar. We engineer each pillar's shapes and metallic covering by optimizing the scattering process for predicting the analyte's concentration. 
\label{fab} 
}
\end{figure*}

\begin{figure*}
\centering
\includegraphics[width=0.99\textwidth]{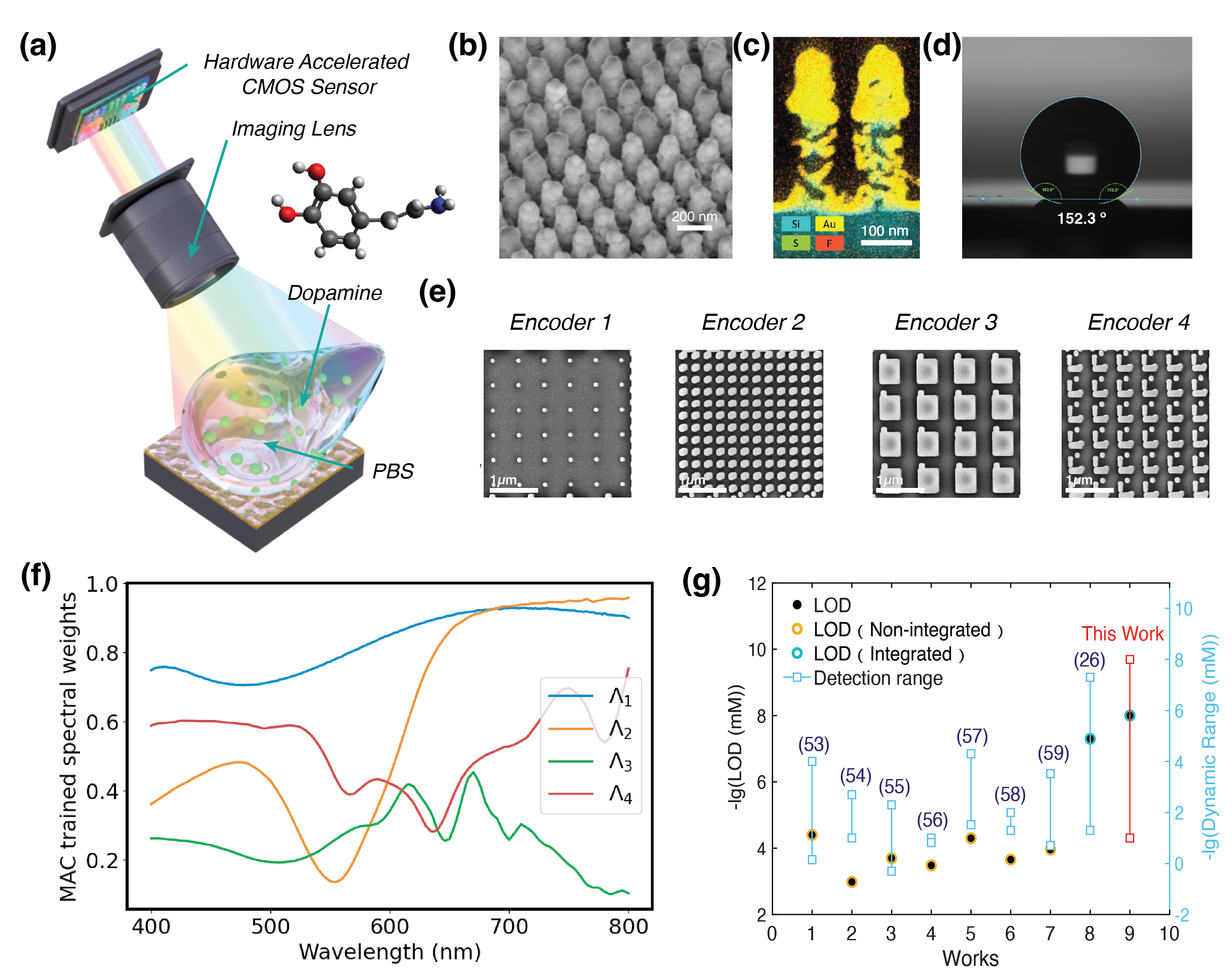}
\caption{\textbf{Experimental results on DA interfered with PBS.} (a) Schematic of the setup for measuring DA in PBS. (b) Tilted SEM and (c) EELS-elemental mapping image of the fabricated optical scattering device. (d) CA image of the scatterer treated by PFDT. (e) SEM image of the first three hardware encoders and (f) trained transmission spectra weights functions.  (g) LOD and dynamic range of DA in PBS compared against the state-of-the-art approaches.
\label{Dopamine_PBS} 
}
\end{figure*}

\begin{figure*}
\centering
\includegraphics[width=0.99\textwidth]{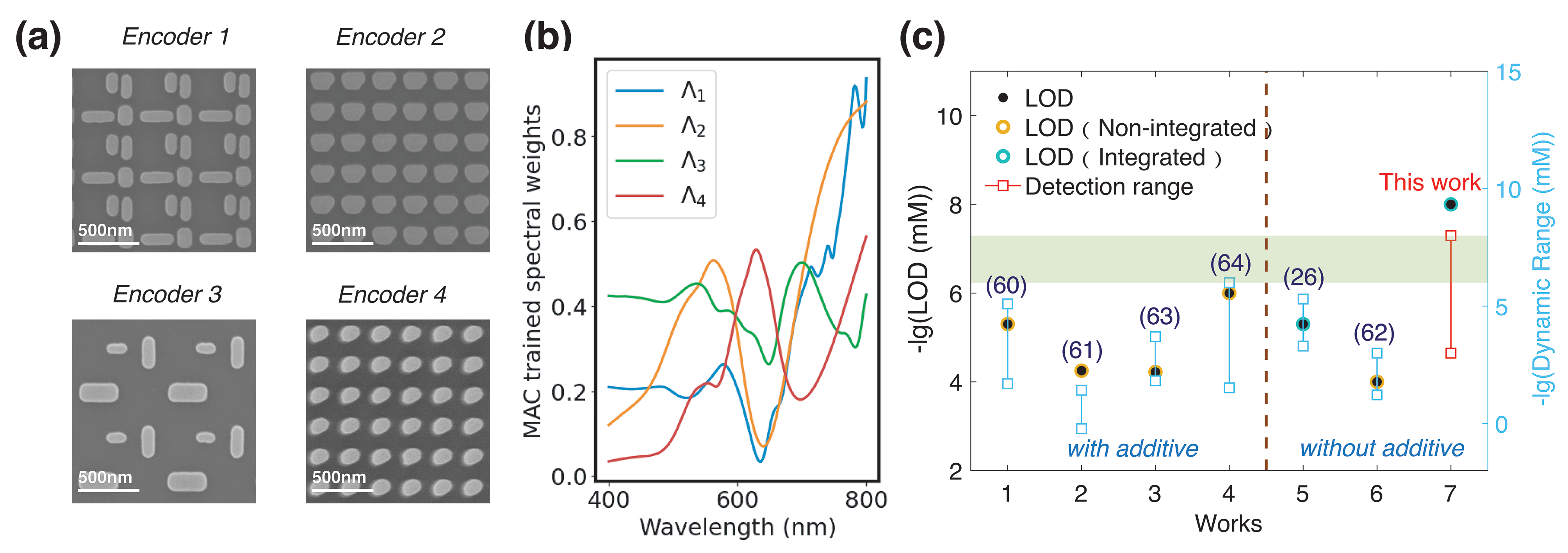}
\caption{\textbf{Detection of DA interfered with UA and AA.} (a) SEM image of first three hardware encoders with corresponding (b) measured trained spectral weights. (c) LOD and dynamic range of DA detection compared against the state-of-the-art. The green rectangle region linked to the right y-axis shows the DA concentration fluctuation range in most human brain disorders.
\label{Dopamine_UA} 
}
\end{figure*}

\noindent Figure~\ref{frame} provides a general scheme of the ultra-sensitive detection platform we implement in this work. Input light with spectral intensity distribution $I_i(\omega)$ impinges on an analyte with unknown concentration $y$, placed inside a suitably designed scattering surface (Fig.~\ref{frame}a). When illuminated by a broadband input $I_i(\omega)$, the surface outputs a scattering response $I_o(\omega,y)$ that could leverage different forms of light-matter interaction, including plasmonic resonances in metallic nanostructures~\cite{lalanne2018light}, Mie or Fano resonance in high-refractive-index dielectrics \cite{kivshar2017meta}, and chemical bonding between the analyte and molecules with high scattering cross section~\cite{lei2020singleatom}.\\
The spectra obtained at different analyte concentrations $y$ compose a training dataset (Fig.~\ref{frame}b). The dataset contains sparse spectral features relating the concentration $y$ with the input beam $I_i(\omega)$ spectrum. The platform uses the spectral dataset to train a hardware encoder $\mathcal{E}$, which extracts spectral features using artificial intelligence hardware (Fig.~\ref{frame}c). The encoder $\mathcal{E}$ follows the methodology of unsupervised dimensionality reduction~\cite{wang2016auto}, and transforms high-dimensional scattered spectra $I_o(\omega,y)$ into a latent vector $\mathbf{z}=\mathcal{E}(I_o(\omega,y))$, representing a low-dimensional feature space of the input spectra.
A decoder model $\mathcal{D}$ employs the information contained in the latent vector $\mathbf{z}$ to reconstruct the input spectra, providing at the output the set of spectra $I'_o(\omega,y)$ as close as possible to the input ones. Training of the  $\mathcal{E}$- $\mathcal{D}$ systems comprises minimizing the mean square error (MSE) loss~\cite{hecht1992theory} between the input and reconstructed spectra: $\mathcal{L}_{rec}=\frac{1}{N}\sum_{y}\| I_{o} (\omega ,y )-I'_{o} (\omega,y) \|$, 
where $N$ represents the total number of spectra samples. As the training converges, the encoder $\mathcal{E}$ provides an information-preserving projection from the spectral scattering responses to the distribution of all possible low-dimensional representations, yielding a latent space encompassing the important sparse features of the input spectra $I_{o} (\omega,y )$. The proposed framework consequently applies a supervised feature selection process in the latent space using explainable AI (XAI)~\cite{lundberg2020local}, which optimizes the number of latent features for predicting the analyte concentration $y$ (arrow under Fig.~\ref{frame}c).\\
We implement the encoder $\mathcal{E}$ in optoelectronic hardware using nanostructured encoder elements with engineered transmission responses in the frequency domain $\Lambda_i(\omega)$ (Fig.~\ref{frame}d) placed over the sensor of a conventional monochrome camera. In this process (Fig.~\ref{frame}d-e), each component $z_i$ in the latent vector $\textbf{z}$ is hardware generated from input light $I_o$ transmitted through a surface $S_i$ with transfer function $\Lambda_i(\omega)$, and then converted into electronic signal in the photo-detector:
\begin{equation}
    \label{proj}
    z_i=\sigma\left[\int_{\omega_0}^{\omega_1} \dd\omega I_o(\omega,y)\Lambda_i(\omega)\right]\equiv \langle I_o(\omega,y),\Lambda_i(\omega)\rangle,
\end{equation}
with $\Delta\omega=\omega_1-\omega_0$ being the sensitivity bandwidth of the detector and $\sigma$ its nonlinear saturable transfer-function~\cite{liu1999saturation}, or an additional post-conversion nonlinearity computed on the electronic stream. Equation \eqref{proj} implements a multiply and accumulate (MAC) operation of neural networks (NN), with the camera conversion function $\sigma$ resembling the corresponding activation function in NN. In the projection process described by Eq.~\eqref{proj}, we inverse design the transmission $\Lambda_i$ with an open-source inverse design software~\cite{makarenko2021robust} to yield the closest representation of $z_i$ in the least-square sense (Fig.~\ref{frame}f). To ensure the solution to this problem exists, we implement $S_i$ with a single metasurface containing a layer of physical universal approximators represented by nanoscale resonators with tailored resonance responses~\cite{makarenko2022realtime}. We arrange the set of $S_1,...S_N$ projecting surfaces into a matrix of sub-pixel arrays projecting on the camera pixels (Fig.~\ref{frame}g). The image readout from the camera acquires the latent vector $\mathbf{z}$ and flattens it into an electric signal, which the regression network processes to predict the analyte concentration $y$~(Fig.~\ref{frame}g).\\
Figure~\ref{design} details the hardware encoder $\mathcal{E}$ feature extraction process, using an input spectral dataset $\textbf{X}=[I_o(\omega,y)_1,I_o(\omega,y)_2,...,I_o(\omega,y)_N]$ obtained from DA dissolved in PBS. In this work, we acquire spectral datasets with a commercial Ocean Optics UV-VIS spectrum analyzer equipped with an integrating sphere. The dataset $\textbf{X}$ comprises $N=240$ reflection spectra equally measured in 8 concentrations varying from $10^{-1}$~mM to $10^{-8}$~mM (Fig.~\ref{design}a). The training and testing portion of the dataset comprises 80\% and 20\% of the total samples, respectively. In this work, we perform dimensionality reduction by principal component analysis (PCA) (Fig.~\ref{design}b). PCA is a widely used technique in spectral material analysis and classification~\cite{wold1987principal,doi:10.1021/jasms.3c00322,FAIRLEY2023100447} that assists in finding the smallest group of uncorrelated and repeated spectral shapes, which linearly combine to reconstruct the dataset $\textbf{X}$. PCA encodings implement a projection camera process with $\sigma(\mathbf{I})=\mathbf{I}$, which maximizes the data processing speed. The PCA eigenvector components represent the MAC weights $\boldsymbol\Lambda(\omega)=[\Lambda(\omega)_1,...,\lambda_M(\omega)]$ for training the hardware accelerator encoder in Fig.~\ref{frame}f. The first $M=8$ components represent the spectral data up to a total of 99.8\% explained variance and result in a dataset reconstruction loss $ \mathcal{L}_{rec} = 1.2\times 10^{-6}$. Figure~\ref{design}b projects the input spectral dataset $\textbf{X}$ along the first three principal components. The principal features provide a new set of coordinates to represent the data with successive refinement, serving as a basis for feature separation and identification. The projected points in Fig.~\ref{design}b show this property quantitatively, separating data distribution patterns in the space where high-concentration and low-concentration data points cluster in different areas.\\
We identify the most useful principal components for analyte concentration prediction using the XAI integrated gradient (IG) method~\cite{sundararajan2017axiomatic}. The IG trains an auxiliary regression network $F$ to predict the analyte concentration (green block in Fig.~\ref{design}c). The network $F$ uses input data comprising the spectral features $\textbf{z}=[z_1,...,z_M]$, with $z_i$ being the projection of the input spectra along the $i$-th principal component $u_i$. The training of $F$ aims to minimize the regression loss $\mathcal{L}_{reg}=||\hat y -y||$ between its prediction output $\hat y=F(\textbf{z})$ and the ground truth concentration $y$ obtained from measurements.
The IG assigns to each input feature $z_i$ an importance score (IS),  computed from:
\begin{equation}
\label{ig}
\text{IS}(z_i) = (z_i - z'_i) \times \int_{\alpha=0}^{1} \frac{\partial F(\textbf{z}' + \alpha (\textbf{z} - \textbf{z}'))}{\partial z_i} \, d\alpha,
\end{equation}
with $\textbf{z}'=\textbf{0}$ representing a 'void sample,' and the integral accumulating the partial derivatives of prediction output $\hat y$ with regard to the input feature $z_i$. The IG method constructs an integral path from $\textbf{z}'$ to $\textbf{z}$ by linearly interpolating between them (Fig.~\ref{design}c). Along this  path, it computes the gradient on inputs $\textbf{z}' + \alpha (\textbf{z} - \textbf{z})'$ via backpropagation, with $\alpha \in [0,1]$ being the interpolation factor.
The integral value in \eqref{ig} is rescaled by $z_i - z_i'$ to ensure that the IS is proportional to the distance between the input and the void sample. The larger the IS deviates from 0, the greater significance $z_i$ carries for predicting the analyte concentrates $y$. 
Figure~\ref{design}d plots the absolute values of the calculated IS and the explained variances of each principal component. While consecutive principal components represent the input dataset with accumulating total variance (Fig.~\ref{design}d dashed line), the IG analysis highlights that the four most informative components for predicting an analyte concentration are in a nonconsecutive order (Fig.~\ref{design}d, red solid lines). Further reducing the PCA components to the four with the highest IS projects $\textbf{X}$ to a feature space with an explained variance of 83.92\% for the input data while carrying a total IS of 86.99\%.
Figure~\ref{integration}a shows the encoder integration on top of a commercial DMM 36AX290-ML board monochrome camera from The Imaging Source. Figure~\ref{integration}b shows a scanning electron microscope (SEM) image of the hardware encoders, with a one-to-one mapping between the camera sensor pixels and the metasurface sub-pixels (see fabrication details in Methods).\\  
Figure~\ref{fab}a illustrates the scattering device implemented in this study (details in the Method section). The system integrates a scattering plasmonic system in a microfluidic cell, comprising a central cylindrical cavity with a diameter of 10~mm, capable of accommodating \SI{1}{\milli\liter} of solution. The device includes two input and output analyte channels, each with a diameter of 0.5 mm, positioned on either side of the cavity (Fig.~\ref{fab}a, solid arrows). Figure~\ref{fab}b shows the schematic of the device's assembly. The bottom scattering layer comprises a nanostructured silicon (Si) wafer decorated with metallic nanoparticles (Fig.~\ref{fab}c). 
We fabricate nanostructured Si using wet-chemistry self-assembly followed by nanostructuring via reactive ion etching (RIE). The RIE process is separately optimized for each analyte, leveraging the available RIE parameters comprising the gas flow of \ce{SF6}, \ce{CHF3}, \ce{O2}, the set pressure (SP), the power of
inductively coupled plasma (ICP), the power of radio frequency (RF), and etching time (ET). The optimization process maximizes the separation of features in the latent space generated by the hardware encoder $\mathcal{E}$ (details in the Methods section). The device's top layer is a hydrophobic multilayer structure, which prevents the analyte from contaminating the scattering substrate, allowing the sensor to be reusable. The present system implements a bottom layer of 1H,1H,2H,2H-Perfluorodecanethiol (PFDT, \ce{CF3(CF2)7CH2CH2SH}), an interlayer of Polydimethylsiloxane (PDMS) and cover glass treated by Perfluorodecyltrichlorosilane (FDTS, \ce{CF3(CF2)5CH2CH2SiCl3}).\\
Figure~\ref{Dopamine_PBS}a shows the light path and the device assembly for measuring DA in PBS, with Figs~\ref{Dopamine_PBS}b-c detailing the nanoscale images of the scattering substrate engineered in the microfluidic cell. 
The SEM image in Fig.~\ref{Dopamine_PBS}b shows that the RIE-optimized nanostructured for DA appears as elongated pillars on a hexagonal lattice with an average period of 200~nm and 535~nm height, characterized by a disordered profile modulation along the transverse $z$ axis. Figure~\ref{Dopamine_PBS}c provides additional details by showing scanning transmission electron microscopy-electron energy loss spectroscopy (STEM-EELS) elemental mapping of a material section comprising two consecutive nanopillars. 
Metallic Au nanoparticles deposited via sputtering accumulate primarily on top of the pillars, providing mushroom-like scattering shapes. Supplementary Figure~S1 shows individual STEM-EELS mapping of each component analyzed, including Si, Au, Sulfur (S), and Fluorine (F). Elements S and F, originating from assembled PFDT molecules, align with the regions where Au appears on the scatterer. 
This feature indicates that assembled PFDT via the liquid-phase growth method contributes to the inhomogeneous surface metal decoration observed in Fig.~\ref{Dopamine_PBS}c.\\ 
Supplementary Figure~S2 uses X-ray Photoelectron Spectroscopy (XPS) to investigate the mechanisms of PFDT  layer deposition over the substrate. 
PFDT-treated scatterer shows an S 2p doublet centered at $162.0 \pm 0.2$ eV (Suppl. Fig.~S2a), indicating the formation of an Au-S bond between the thiol group and the gold substrate in agreement with the previous report~\cite{hamoudi2010self}. The C 1s element shows intensity peaks at 283.3, 289.9, and 292.2 eV (Suppl. Fig. S2b), corresponding to the formation of \ce{-CH2-}, \ce{-CF2-}, and \ce{-CF3-} bonds, respectively~\cite{dass2008gold}. These results demonstrate that PFDT distributes on the scatterer through chemical bonding without physical absorption. Chemical bonding guarantees that PFDT elements are not dissolved in the microfluidic cell at the moment of the measurement, avoiding contamination with the analyte.\\ 
Figure~\ref{Dopamine_PBS}d employs contact angle (CA) measurements to quantify the hydrophobicity of PFDT-treated scatterers. 
The CA observed lies above $152.3 \pm 0.6^\circ$ with the relative standard deviation (RSD) of 0.36\% computed from 10 different measurements, showing super-hydrophobic behavior (Suppl.  Fig.~S3). 
The CA observed for the FDTS-treated cover glass is $102.2 \pm 0.5^\circ$ (Suppl.  Fig.~S4), slightly lower than PFDT but still well above the hydrophobic threshold of 90$^\circ$.\\ 
Figure~\ref{Dopamine_PBS}e shows the morphology of the nanostructured metasurfaces representing the four XAI-selected principal components trained by the procedure of Fig.~\ref{design} in the case of DA analyte in PBS.  
Figure~\ref{Dopamine_PBS}f illustrates the measured spectral responses $\Gamma_i(\omega)$ ($i=1,2,3,4$) of the trained MAC encoders.
Figure~\ref{Dopamine_PBS}g (solid red color) summarizes the detection results of the proposed platform and compares them against the state-of-the-art (solid blue lines)~\cite{lei2020singleatom,wu2015dopamine,roychoudhury2016dopamine,yuan2018highly,xu2018sensitive,sun2014gold,yang2017using,zhao2019situ}. 
The dynamic range achievable with this technique spans eight orders of magnitudes, from $\mathrm{10^{-8}}$ mM to $\mathrm{10^{-1}}$ mM, while the $R^2$ score estimated from the test dataset reaches 0.9898. This performance represents an enhancement of the limit of detection (LOD) of 80.0\% and three orders of magnitude compared to the best integrated and non-integrated state-of-the-art, respectively. It also provides the broadest dynamic range, surpassing the best \ce{Mn-MoS2}/PGS electrochemical device by two orders of magnitude~\cite{lei2020singleatom}.\\  
Figure~\ref{Dopamine_UA} summarizes detection results for measuring DA in a PBS solution with the interference of AA and UA. 
Addressing performances with these interference media is necessary for the reliability of DA reading under physiological conditions and crucial for point-of-care diagnostics~\cite{lei2020singleatom}. 
Figure~\ref{Dopamine_UA}a shows the SEM image of the nanostructures from the four XAI-selected light encoders that extract the required features for this type of analyte in the presence of AA and UA. It also depicts the corresponding trained MAC spectral weights in Fig.~\ref{Dopamine_UA}b. Similarly to the analysis of PBS, we collected 180 spectra data reflected from samples with different concentrations and split the training/test dataset by 4:1. We further extracted four PCA components carrying 87.12\% of IS and 96.40\% of the total variance of the training dataset. Fig.~\ref{Dopamine_UA}c shows the detection results compared to the state-of-art by using a hardware encoder implementing the four selected PCA components.
We here obtain LOD of $\mathrm{10^{-8}}$~mM and a dynamic range spanning from $\mathrm{10^{-8}}$ to $\mathrm{10^{-3}}$~mM. Analysis of the test dataset reports an $R^2$ score of 0.9926. Comparable to available technologies~\cite{cui2017electrochemical,demirkan2020palladium,zhang2020electrochemical,kunpatee2020simultaneous,ji2020carbonized,lei2020singleatom}, the integrated technique presented succeeds in overcoming the threshold limit range from $10^{-8}$~mM to $10^{-6}$~mM (green rectangle region) required to monitoring DA fluctuations in neurological disorders~\cite{sharma2019dopamine,shine2019dopamine,howes2024schizophrenia}.

\section*{Discussions}
This work implements a hardware-accelerated ultra-sensitive platform technology to detect biochemical analytes in real time. The proposed technology integrates conventional camera sensors and does not require extra additives, avoiding traditional impurity treatment requiring between tens of minutes to several hours~\cite{xie2022label} that make continuous motoring impossible.  
The entire manufacturing process for a single detection system is within 1000~USD, evaluated with clean room actual costs. This figure represents 1/10 of the price of a state-of-the-art electrochemical workstation, which also requires a significantly larger footprint than a portable integrated camera.
By reaching the threshold of DA detection at $\mathrm{10^{-8}}$ mM, the proposed technology can enable real-time tracking of addiction-related DA dynamics~\cite{kasai2021spine}, and early diagnosis of Parkinson's disease, facilitating the study of DA's influence on emotional regulation~\cite{schuster2022dopaminergic,zhang2022mining}, and neurotoxic effects~\cite{bido2021microglia}.
The universal design of this sensing platform combining trained optical accelerators with software neural networks can extend this study on a wide range of additional analytes relevant to medical research, including serotonin and norepinephrine for studying depression and anxiety~\cite{bangasser2021sex}, glutamate and gamma-aminobutyric acid for understanding neurodegenerative diseases like Alzheimer's and epilepsy~\cite{cheung2022physiological,barbour2024seizures}, and cytokines indicators for cancer and immune system disorders~\cite{propper2022harnessing}.

\section*{Methods}
\subsection{Materials and reagents}
We purchase polystyrene (PS) latex particles (PS200NM, 10\% w/v) from MAGSPHERE Inc.. We order Sylgard\textsuperscript{TM} 184 from Electron Microscopy Sciences. We acquire ethanol, toluene, and sodium dodecyl sulfate solutions (SDS, 10\% in $\mathrm{H_2O}$), PFDT, PBS, DA, UA and AA from Sigma-Aldrich Inc. We obtain Si wafers from University Wafer Inc.. We purchase the artificial sweat BZ320 from Biochemazone\textsuperscript{TM}. We use deionized water (Milli-Q\textregistered~Advantage A10) with a resistance of \SI{18}{\mega\ohm}$\cdot$cm in all experiments. 

\subsection{Fabrication of the scattering surface}
We clean the Si surface with isopropanol, acetone, and ultra-pure water and then dry it with nitrogen gas. Next, we prepare the PS solutions by mixing the ethanol and water with a volume ratio of 1 : 1 and make these latex particles disperse in the solutions homogeneously by ultrasonic process for 30 min. We then perform self-assembly of PS on the water with the assistance of SDS. After the air-drying process, we etch the sample in the RIE instrument ((Oxford Instruments, Plasmalab 100 - ICP 380)). We use an oxygen plasma process (JST Manufacturing, Inc., ST0122A0) to remove the residual PS. We finally deposit 80~nm gold film on the structured Si substrate in the sputtering process (Equipment Support Company Ltd, ESCRD4), followed by the annealing process (Qualflow Them/Jipelec, JETFIRST) at 200 $\mathrm{^oC}$ for 5 min.

\subsection{Assembly of the scattering device}
We perform an oxygen plasma process on the PDMS layer for 1 minute. We then assemble it with PFDT decorated scatterer and FDTS modified cover glass. After stabilizing the structure for 2 days, the cell is ready for experiments.

\subsection{Fabrication of the subpixel encoders}
Using plasma-enhanced chemical vapor deposition, we deposit amorphous silicon on a fused silica substrate. We consequently spin coat and thermal-process the AR-P 6200.09 (Allresist GmbH) positive electron beam resist and a conductive polymer AR-PC 5090.02 layer on the substrate. We employ a JEOL JBX-6300FS electro-beam lithography system to write the nanostructure pattern, followed by a resist development process. We then deposit a 22 nm layer of chromium on the substrate to form a protective layer in the shape of the patterns to be unetched. We use a RIE system to expose the patterns of silicon. As a final step, we remove the chromium protection mask with perchloric acid and ceric ammonium nitrate solution (TechniEtch Cr01 from MicroChemicals).

\subsection{Physical characterization}
We collect the SEM images on Carl Zeiss Merlin field-emission SEM with an In-Lens detector. We collect the high-angle annular dark field scanning transmission electron microscopy (HAADF-STEM) images and STEM electron energy-loss spectroscopy (STEM-EELS) mapping of the FIB-prepared specimen with dual aberration-corrected TEM (Titan Themis Z, FEI) at the operating voltage of 300~kV. We perform the XPS measurements on the Kratos AXIS Supra system.

\subsection{Data availability.} Source data are provided with this paper.

\subsection{Code availability.} The code used in 
this work is available at   \url{https://github.com/QizhouW/dopamine_sensing}.

\bibliography{paper}

\clearpage

\subsection{Acknowledgements.} We conducted the hardware encoder's design using computing resources provided by KAUST Supercomputing Core Lab. We carried out the hardware encoder's fabrication, scatterer fabrication, scanning electron microscopy, and transmission electron microscopy analysis at KAUST's Nanofabrication, Imaging, and Characterization Core Lab, respectively. We thank KAUST's Core Labs staff for their help in the design and fabrication processes.

\subsection{Author Contributions.} N. Li, Q. Wang and A. Fratalocchi initiated, managed and planned the overall project. N. L. and Z. He completed the fabrication process. Q. Wang finished the machine learning and related theory. A. B. Lopez helped calibrate the spectrometer. N. Li and Q. Wang contributed to this work equally. All authors contributed to data analysis and manuscript refinement and preparation.

\subsection{Correspondence} Correspondence and requests for materials
should be addressed to ning.li@kaust.edu.sa and andrea.fratalocchi@kaust.edu.sa.

\subsection{Competing interests.} The authors declare no competing interests.

\clearpage

\clearpage

\end{document}


\maketitle

\begin{affiliations}
 \item PRIMALIGHT, Faculty of Electrical Engineering, King Abdullah University of Science and Technology (KAUST), Thuwal 23955-6900, Saudi Arabia.\\ 
 \textsuperscript{\textdagger}first authors with equal contribution
\end{affiliations}

\begin{figure*}
\centering
\includegraphics[width=0.5\textwidth]{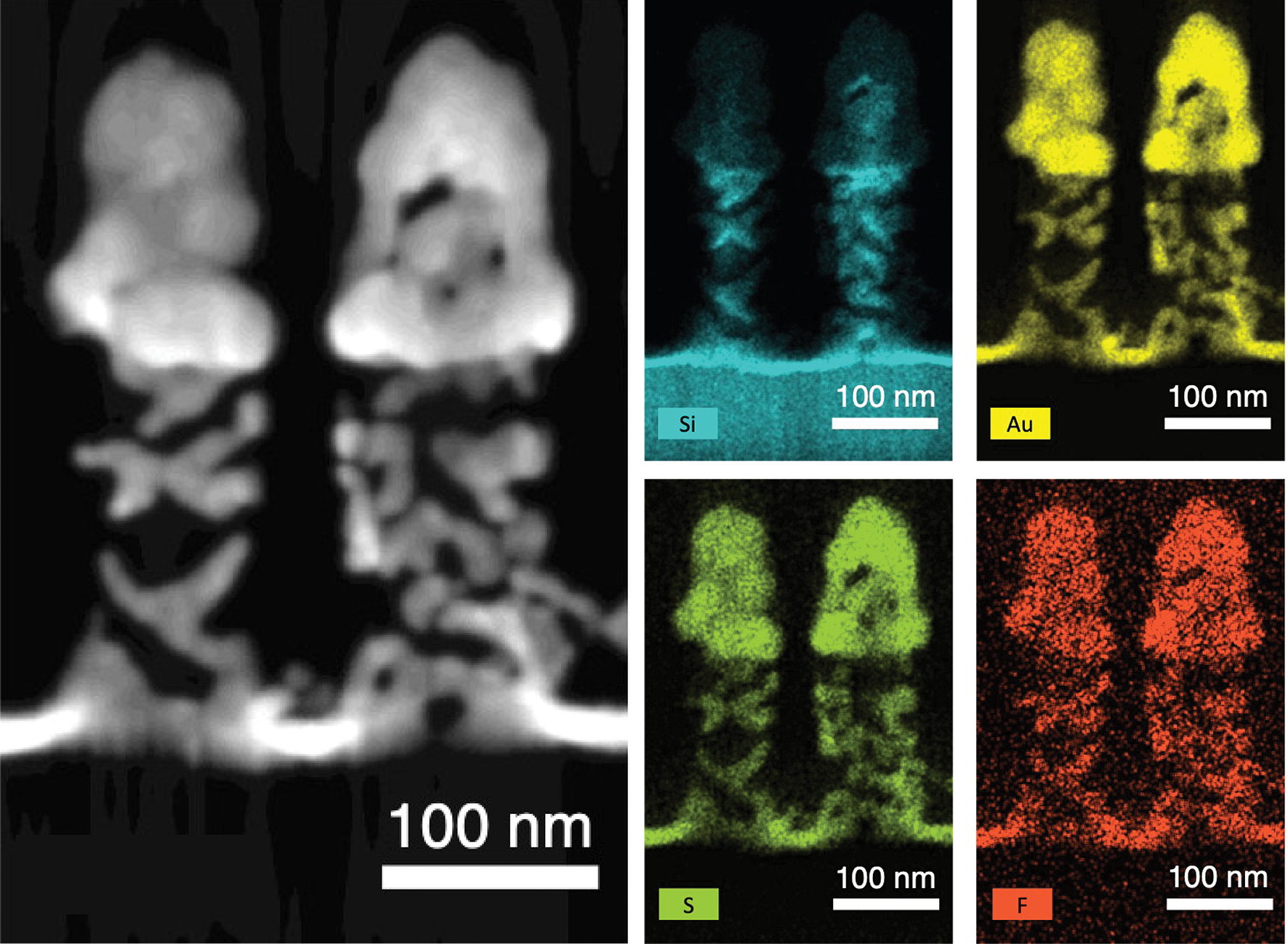}
\caption{Cross-sectional HAADF-STEM image and EDS elemental mapping of the Si, Au, S and F.
\label{Element_mapping} 
}
\end{figure*}

\begin{figure*}
\centering
\includegraphics[width=0.6\textwidth]{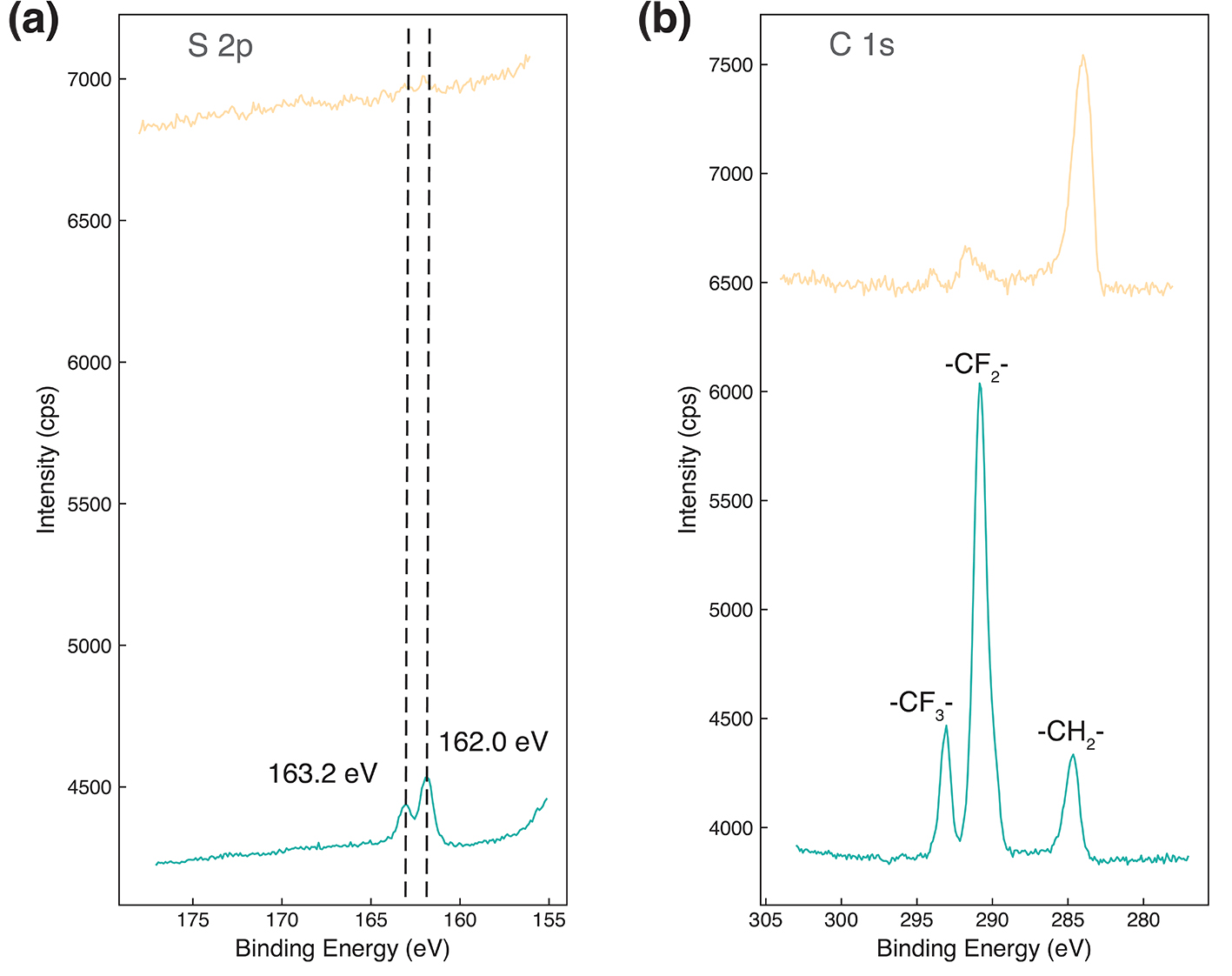}
\caption{XPS of scatterer (orange curve) and PFDT decorated scatterer (blue curve) : (a) S 2p, and (b) C 1s.
\label{XPS} 
}
\end{figure*}

\begin{figure*}
\centering
\includegraphics[width=0.99\textwidth]{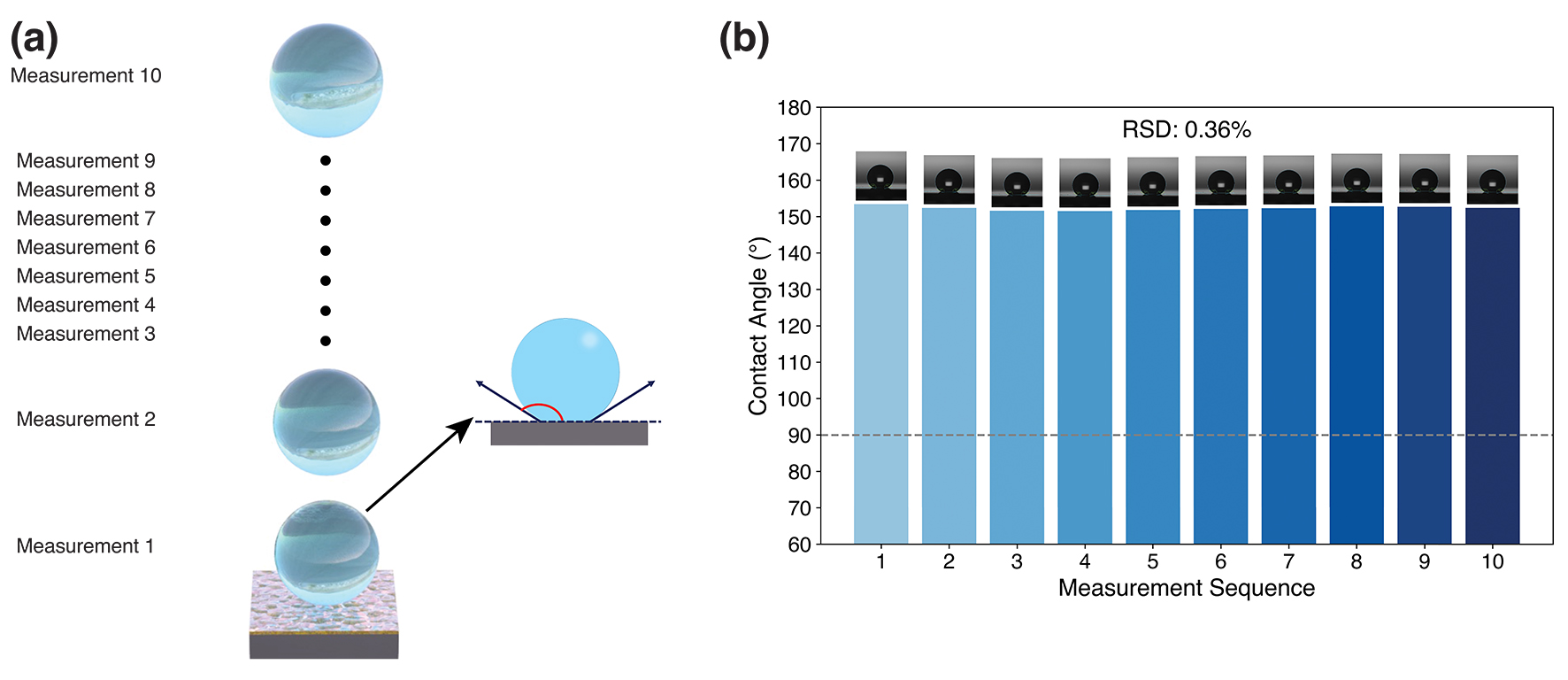}
\caption{(a) Schematic of the measuring CA on the PFDT decorated scatterer at the same point for 10 times. (b) Statistics of 10 times CA measured on the PFDT decorated scatterer.
\label{Ten_measurement} 
}
\end{figure*}

\begin{figure*}
\centering
\includegraphics[width=0.5\textwidth]{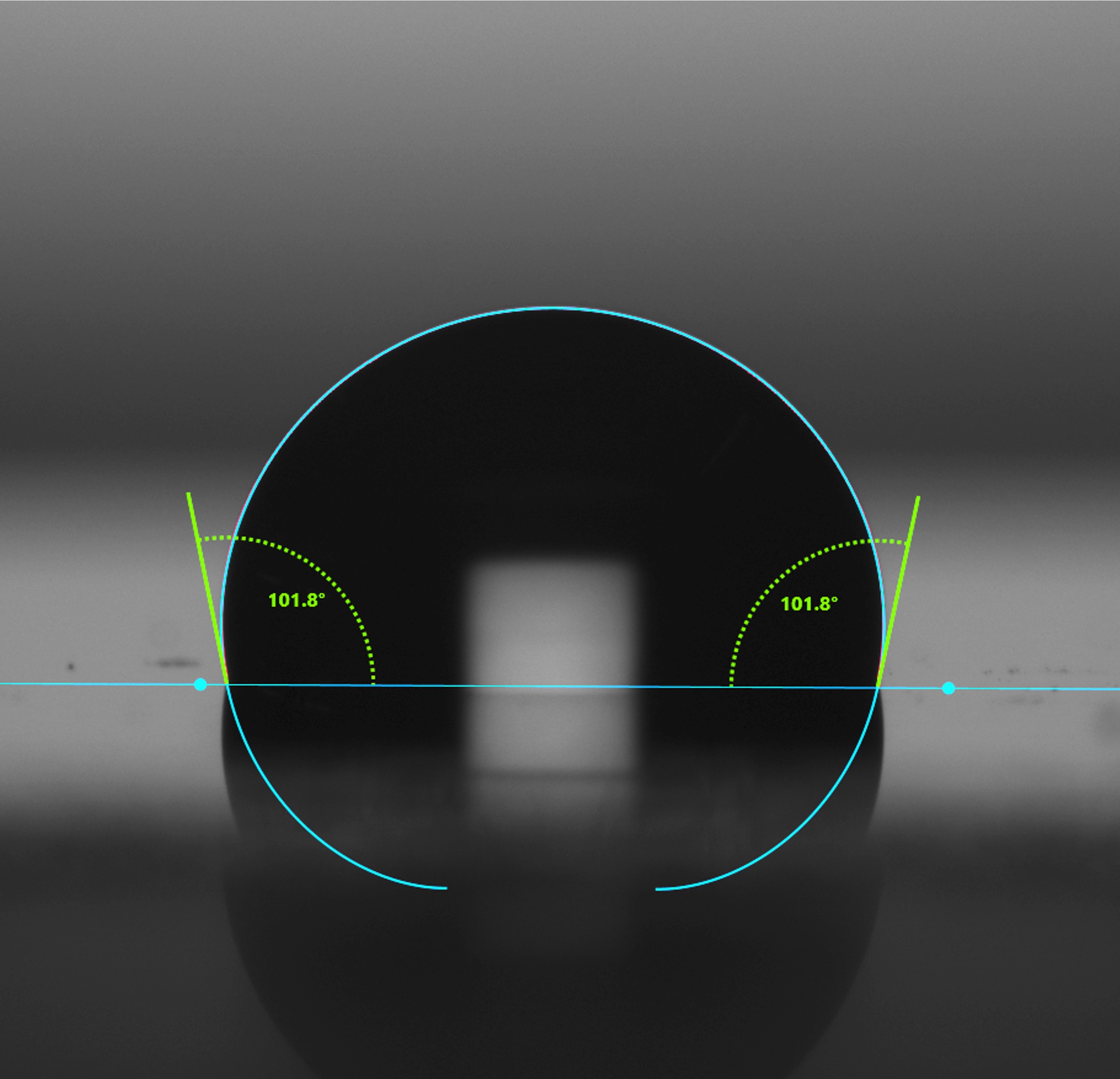}
\caption{CA image of the cover glass treated by FDTS.
\label{CA_Glass} 
}
\end{figure*}